\documentclass[twocolumn,showpacs,preprintnumbers,amsmath,amssymb]{revtex4}

\usepackage{graphicx}
\usepackage{dcolumn}

\newcommand{\n}{\nonumber}
\newcommand{\mpc}{\mbox{\ $h^{-1}$~Mpc}}

\begin{document}

\title{On the onset of cosmological backreaction}

\author{Nan Li\footnote{Electronic address: li@physik.uni-bielefeld.de}}
\author{Dominik J. Schwarz\footnote{Electronic address: dschwarz@physik.uni-bielefeld.de}}

\affiliation{Fakult\"at f\"ur Physik, Universit\"at Bielefeld,
Universit\"atsstra\ss e 25, D-33615 Bielefeld, Germany}

\begin{abstract}
Cosmological backreaction has been suggested as an explanation of
dark energy and is heavily disputed since. We combine cosmological
perturbation theory with Buchert's non-perturbative framework,
calculate the relevant averaged observables up to second order in
the comoving synchronous gauge, and discuss their gauge dependence.
With the help of an integrability condition, the leading second
order contributions follow from the first order calculation.

We focus on the onset of cosmological backreaction, as a
perturbative treatment is necessarily restricted to the era when the
effect is still small. We demonstrate that the leading contributions
to all averaged physical observables are completely specified on the
boundary of the averaged domain. For any finite domain these surface
terms are nonzero in general and thus backreaction is for real. We
map the backreaction effect on an effectively homogeneous and
isotropic fluid. The generic effective equation of state is not only
time dependent, but also depends on scale.
\end{abstract}


\maketitle

\section{Introduction} \label{sec:intro}

The accelerated expansion of the Universe has now been confirmed by
various
observations~\cite{Astier:2005qq,Riess:2006fw,Conley:2006qb}. To
understand this mysterious phenomenon, different explanations have
been suggested, e.g., dark energy in the form of a quintessence
field or a modification of gravity on large scales. Among the many
proposals, a cosmological constant (or vacuum energy) in the context
of the inflationary $\Lambda$ cold dark matter ($\Lambda$CDM) model
seems to be the most attractive one, as it is simple and provides a
good fit to all cosmological data currently available.

However, the observational evidence is based on the assumption of a
homogeneous and isotropic universe (the
Friedmann-Lema\^{i}tre-Robertson-Walker (FLRW) model). But, spatial
homogeneity and isotropy are rather rough approximations for the
Universe, only valid on scales larger than $\sim 100
\mpc$~\cite{Hogg:2004vw,Joyce:2005rg}. Thus, before assuming dark
energy to be a component of the Universe, it is worthy to
investigate the effects of local inhomogeneities and anisotropies.
On the largest scales, the deviations from homogeneity and isotropy
are tiny and a description in terms of linear cosmological
perturbations is very well justified. However, when the matter and
metric fluctuations enter the nonlinear regime (currently the
largest observed non-linear objects extend to the $100 \mpc$ scale,
among the most prominent the Shapely supercluster and the Sloan
great wall), the problem of averages~\cite{Ellis,Zalaletdinov:1992cg} 
shows up, and as a consequence the
observed cosmological parameters might actually differ from the
actual parameters of the underlying cosmology.

Let us pick the most important cosmological parameter, the Hubble
expansion rate~\cite{Freedman:2000cf,Sandage:2006cv}
to discuss this issue in some detail. The idealized
measurement of the Hubble expansion rate proceeds as follows. Take a
set of $N$ standard candles (in reality supernovae of type Ia) that
sample a local (i.e., objects at redshift $z \ll 1$) physical volume
$V$ homogeneously. Measure their luminosity distances $d_i$ (via
magnitudes) and recession velocities $v_i = c z_i$ and take the
average~\cite{Tully:2007ue}
\begin{equation}
H_0 \equiv \frac{1}{N}\sum_{i=1}^N \frac{v_i}{d_i}.\n
\end{equation}
In the limit of a very big sample ($N \to \infty$), this turns into
a volume average
\begin{equation}
H_0 = \frac{1}{V}\int \frac{v}{d} {\rm d}V.\n
\end{equation}
In the second step, we neglect the effect of the light cone, but for
$z\ll 1$ the spatial average is a good approximation for an average
over the past light cone, because the expansion rate of the Universe
is not changing significantly at time scales much shorter than the
Hubble time.

On the other hand, we have a theoretical object that we call the
expansion rate, defined as $H_0^{\rm th} \equiv \dot{a}/a$, where
$a$ denotes the scale factor of the background model. The issue in
the averaging problem now is to establish the connection between
$H_0$ and $H_0^{\rm th}$. In linear theory, by construction, the
average $H_0$ and the background $H_0^{\rm th}$ agree if the volume
$V$ becomes large enough. However, due to the nonlinearity of the
Einstein equations, cosmological perturbations can affect the
evolution of the average, which we often identify with the
``background'' Universe. This is the so-called backreaction
mechanism~\cite{Ellis,Zalaletdinov:1992cg,Mars:1997jy,Buchert:1999er,
Buchert:1999mc,Buchert:2001sa,Geshnizjani:2002wp,Zimdahl:2000zm,
Schwarz:2002ba,Rasanen:2003fy,Aghababaie:2003wz,
Kolb:2004am,Burgess:2004yq,Kolb:2005da,Coley:2005ei,
Rasanen:2006kp,Parry:2006uu,Coley:2006xu,Coley:2006kp,Buchert:2006rq,Buchert:2007ab}.

As in the example of the Hubble constant, we typically measure the
value of a physical observable by taking an average of the
observable in a domain of space-time. Other examples are number
counts, correlation functions, power spectra, etc. Consequently, a
comparison of the theory with observation needs to utilize averaged
quantities.

The influence of cosmological perturbations on the expansion of the
Universe shows up in many aspects, e.g., for an underdense patch of
the Universe the local expansion rate is naturally larger than its
average. The task is to find the evolution equations for the
averaged observables and thus the effective equation of state of the
Universe.

If this effective equation of state $w_{\rm eff} \equiv p_{\rm
eff}/\rho_{\rm eff}<-1/3$, where $\rho_{\rm eff}$ and $p_{\rm eff}$
are the effective energy density and pressure, the expansion of the
averaged Universe would accelerate. So, if backreaction (i.e., averaging)
would give rise to a negative effective pressure, we might be able
to explain the observed acceleration of the Hubble expansion.

At the same time such a mechanism might be able to resolve the
coincidence problem: why does the onset of acceleration happen
around the present time? The answer of cosmological backreaction
could be that originally tiny perturbations grow with time, and lead
to the formation of (weakly) nonlinear structures at the scales that
we use to fix the cosmological parameters. Thus, any observation
that is based on physics in the (weakly or strongly) nonlinear
regime might be influenced by the backreaction effect, especially
the determination of $H_0$ (based on local measurements), the Hubble
diagrams from supernovae Type Ia (need data at low redshift and thus
local information), the integrated Sachs-Wolfe effect, which can
be mistaken for the nonlinear Rees-Sciama effect, and others.

The discussion above does not imply that we have solved the puzzle
of dark energy by the backreaction mechanism. It could well be that
the backreaction effect is tiny and the nature of dark energy is
indeed a constant vacuum energy density. Nevertheless, the
backreaction effect is of interest, because with increasing
experimental precision, e.g., for the cosmic microwave background
(CMB) effects of order $10^{-7}$ are measured \cite{Spergel:2006hy},
and we must take the backreaction effect into account seriously.

Recent research on the backreaction mechanism explored two
directions. One is to study the properties of the averaged physical
quantities in the perturbed Universe. In Buchert's
work~\cite{Buchert:1999er,Buchert:1999mc,Buchert:2001sa}
(see~\cite{Buchert:2006rq,Buchert:2007ab} for a recent review), the
averaged Einstein equations were derived in the synchronous
coordinates with two fluctuation terms, the kinematical backreaction
term $\langle Q\rangle_D$ and the averaged spatial curvature
$\langle R \rangle_D$. The behavior of the perturbed Universe thus
depends on the properties of these averaged terms. In Buchert's
approach, the average is taken over a physically comoving spatial
volume. This seems to be the appropriate procedure for observations
in our local neighborhood, such as for galaxy redshift surveys,
which are currently limited to small redshifts $(z<1)$. However, for
$z>1$, spatial averaging does not reflect actual observations, as we
observe the past light cone and not a spatial hypersurface. We
therefore consider spatial averaging as a first reasonable approach
to the cosmological backreaction problem.

The second direction is to use cosmological perturbation theory to
study the evolution of the perturbed Universe, such as
\cite{Wetterich:2001kr,Rasanen:2004js,Rasanen:2005zy,Hirata:2005ei,
Kolb:2004am,Kolb:2005da,Kolb:2005me,Notari:2005xk,Barausse:2005nf,
Ishibashi:2005sj} and references therein. All these works discussed
the possibility to explain the accelerated expansion of the Universe
as the result of structure formation, without introducing dark
energy into the Einstein equations. For example,
in~\cite{Kolb:2004am}, the Hubble expansion rate was calculated to
second order in a dust (i.e., matter-dominated) Universe.

Recent review and criticism on the backreaction mechanism can be
found in~\cite{Rasanen:2006kp,Buchert:2007ab} and references
therein.

Here, we synthesize these two lines of research. Because doing
perturbative calculations without averaging, we cannot obtain the
global property of the Universe. Whereas, averaging without
perturbative calculations, we cannot get quantitative information of
the Universe. Therefore, in this paper, we calculate the averaged
physical quantities in the cosmological perturbation theory to
second order, but without the need to use the metric perturbations
of second order. In contrast to previous studies in the literature,
we do not aim at calculating ensemble means or variances of spatial
averages. It seems to us that the ensemble means of spatially
averaged quantities are of limited interest for the interpretation
of actual observations. Our interest must be to quantify the
backreaction effect in the Milky Way's neighborhood (the local
spatial domain of $\sim (100~\mpc)^3$, i.e., the domain used to
measure $H_0$). We aim at predicting the amount of backreaction
based on a measurement of the distribution of matter density within
that domain. As we show in this work, the knowledge of the peculiar
gravitational potential on the boundary of a physically comoving
domain at some initial time allows us to predict the time evolution
of the spatially averaged quantities (as long as the effect is
small).

Our paper is organized as follows. In Section 2, we introduce the
concepts of the expansion, shear and rotation of the Universe, and
use the ADM decomposition to rewrite the Einstein equations in terms
of these quantities. Following the averaging procedure of Buchert,
we arrive at the averaged Einstein equations (Buchert
equations~\cite{Buchert:1999er}) for an irrotational dust universe
and an integrability condition in Section 3. The integrability
condition provides a consistency relation for the two backreaction
terms $\langle Q \rangle_D$ and $\langle R \rangle_D$. With Section
4 we turn to cosmological perturbation theory, and solve for the
first order metric perturbations $\Psi$ and $\chi$ in the comoving
synchronous gauge. In Sections 5 and 6, we calculate the
backreaction terms $\langle Q \rangle_D$ and $\langle R \rangle_D$,
the averaged expansion rate $\langle \theta \rangle_D$, the averaged
energy density $\langle \rho \rangle_D$, the effective equation of
state $w_{\rm eff}$ and the square of the effective speed of sound
$c_{\rm eff}^2$ to first and second orders, respectively. Finally,
we demonstrate in Section 7 that the backreaction terms $\langle Q
\rangle_D$ and $\langle R \rangle_D$, as well as $w_{\rm eff}$ and
$c_{\rm eff}^2$ are gauge independent. Conclusions and discussions
are provided in Section 8.

In the following, the Greek indices run from 0 to 3 and the Latin
indices from 1 to 3, and we use units with $c=1$.

\section{Kinematics and dynamics of the expanding Universe} \label{sec:kine}

The standard FLRW model is based on the assumption of spatial
homogeneity and isotropy of the Universe. However, these assumptions
are not valid (even approximately) at the scales on which structure
formation happens, i.e., the scales smaller with respect to the
Hubble radius and sufficiently long after the matter-radiation
equality. So, necessarily one must consider not only the expansion,
but also the shear and rotation of the Universe in order to
understand its kinematics thoroughly.

\subsection{Expansion, shear and rotation} \label{sec:rse}

To describe the kinematics of the Universe, we need to calculate the
gradient field of the 4-velocity
$u^{\mu}\equiv\mbox{d}x^{\mu}/\mbox{d}\tau$ of comoving observers,
where $\tau$ is their proper time. We introduce the projection onto
the spatial hypersurface defined by the comoving observers
$h^{\mu}_{\nu}\equiv \delta^{\mu}_{\nu}+u^{\mu}u_{\nu}$. Thus, the
components of the gradient field of the 4-velocity define the
expansion tensor
\begin{eqnarray}
\theta_{\mu\nu} \equiv u_{\mu;\nu}=
h^{\alpha}_{\mu}h^{\beta}_{\nu}u_{\alpha;\beta}=
\frac{1}{3}h_{\mu\nu}\theta+\sigma_{\mu\nu}+\omega_{\mu\nu},\n
\end{eqnarray}
where $\theta\equiv u^{\lambda}_{;\lambda}$, $\sigma_{\mu\nu}\equiv
h^{\alpha}_{\mu}h^{\beta}_{\nu}(u_{(\alpha;\beta)}
-\frac{1}{3}h_{\alpha\beta}u^{\lambda}_{;\lambda})$, and
$\omega_{\mu\nu}\equiv
h^{\alpha}_{\mu}h^{\beta}_{\nu}u_{[\alpha;\beta]}$ are the expansion
scalar, shear tensor and rotation tensor, respectively.

In the following, we restrict our attention to an irrotational
universe, i.e., $\omega_{\mu\nu} = 0$. Neglecting rotations seems to
be a reasonable assumption in the context of inflationary cosmology,
as there exist no seeds for vector perturbations and the
conservation of angular momentum also implies that only nonlinear
effects could lead to a generation of rotation.

The metric of the inhomogeneous and anisotropic Universe may be
expressed in terms of synchronous coordinates
\begin{eqnarray}
\mbox{d}s^2 = -\mbox{d}t^2 + g_{ij}(t,{\bf
x})\mbox{d}x^i\mbox{d}x^j,\n
\end{eqnarray}
where $t$ is the cosmic time, and $\bf x$ denotes the spatial
coordinates. The corresponding nontrivial Christoffel symbols are
\begin{eqnarray}
&&\Gamma^0_{ij}=\frac{1}{2}g_{ij,0}, \quad
\Gamma^i_{0j}=\frac{1}{2}g^{ik}g_{kj,0}, \n\\
&&\Gamma^i_{jk}=\frac{1}{2}g^{il}(g_{jl,k}+g_{lk,j}-g_{jk,l}).\n
\end{eqnarray}
For an irrotational universe it is possible to use comoving
synchronous coordinates (the observer is at rest with respect to the
cosmic medium), i.e., $u^{\mu}=(1,\bf{0})$ and $h^i_j = \delta^i_j$.
Thus, the  nontrivial components of the shear tensor are
\begin{eqnarray}
&&\sigma_{ij}=\theta_{ij}-\frac{1}{3}g_{ij}\theta, \label{sigma}
\end{eqnarray}
and we define the shear scalar as
\begin{eqnarray}
&&\sigma^2\equiv \frac{1}{2}\sigma^{\mu}_{\nu}\sigma^{\nu}_{\mu}
=\frac{1}{2}\left(\theta^i_j\theta^j_i-\frac{1}{3}\theta^2\right).\label{sigma2}
\end{eqnarray}
Furthermore, the nontrivial components of the expansion tensor
become
\begin{eqnarray}
\theta_{ij}=\Gamma^0_{ij}=\frac{1}{2}g_{ij,0}, \quad \theta=
\theta^i_{i} = \frac{1}{2}g^{ij}g_{ij,0}.\label{theta}
\end{eqnarray}
From (\ref{theta}), we have $\theta = \dot{J}/J$, where
$^{^{\textbf{.}}}$ is the derivative with respect to the cosmic
time, $J\equiv \sqrt{\mbox{det}g_{ij}}$ and $\mbox{det}g_{ij}$
denotes the determinant of the metric. Thus,
\begin{eqnarray}
\dot{J}=\theta J.\label{jtheta}
\end{eqnarray}

\subsection{ADM decomposition} \label{sec:adm}

Having obtained all the quantities to describe the kinematics of the
expanding Universe, we turn to its dynamics. For the dust Universe
(in the comoving synchronous coordinates), the only nontrivial
component of the energy-momentum tensor is $T^0_0 = - \rho$, the
energy density of dust.

According to Arnowitt, Deser, and Misner~\cite{ADM}, the Einstein
equations $G_{\mu\nu}=8\pi GT_{\mu\nu}$ in the present
situation can be decomposed into: \\
the energy constraint
\begin{eqnarray}
R+\theta^2-\theta^i_j\theta^j_i=16\pi G\rho,\label{adm00}
\end{eqnarray}
the momentum constraint
\begin{eqnarray}
\theta^j_{i;j}=\theta_{,i},\label{adm0i}
\end{eqnarray}
and the evolution equation
\begin{eqnarray}
\theta^i_{j,0}=-\theta\theta^i_j-R^i_j+4\pi
G\rho\delta^i_j,\label{admij}
\end{eqnarray}
where $R^i_j$ denotes the spatial Ricci tensor and $R \equiv R^i_i$
is the spatial Ricci scalar. Combining the trace of (\ref{admij})
with (\ref{adm00}) and (\ref{sigma2}) leads to the Raychaudhuri
equation~\cite{Raychaudhuri:1953yv}, which links the expansion and
shear scalars together,
\begin{eqnarray}
\dot{\theta}=-\frac{1}{3}\theta^2-2\sigma^2-4\pi G\rho.\label{ray}
\end{eqnarray}

So far, we have not made any approximations apart from neglecting
rotation and restricting the matter to dust. These equations are
satisfied at any point in space-time. However, our observations do
not allow us to measure all the data that would be necessary to put
a well posed Cauchy problem, but realistic observations deliver
averaged quantities. In the next section we discuss the averaged
properties of these equations, and in Sections 5 and 6 we use
cosmological perturbation theory to evaluate the averaged
observables to first and second orders, respectively.

\section{Dynamics of finite domains} \label{sec:buchert}

In the last section, we set up the local dynamical equations for a
general irrotational dust universe, but realistic observations
provide us with averaged quantities. We follow the averaging
procedure by Buchert~\cite{Buchert:1999er}, and obtain the averaged
equations of motion for the irrotational dust Universe.

A particular choice in Buchert's formalism is the focus on comoving
domains. As long as we neglect the difference between cold dark
matter and baryons, the comoving synchronous coordinate system is
uniquely defined. From the theoretical point of view, this appears
to be the most natural set-up, given that we work in synchronous
comoving coordinates. We argue in Sec. 7, that there are convincing
reasons to pick this particular coordinate system, to be the one
that is best adapted to the actual observational situation. However,
we should keep in mind that for observational purposes it is highly
nontrivial to identify comoving domains. Nevertheless, we think that
comoving domains are a reasonable approximation as the effect from
peculiar motion is small compared to the Hubble expansion at scales
beyond $100 \mpc$.

\subsection{Averaging procedure} \label{sec:averaging}

The spatial average of an observable $O(t,\bf x)$ in a physically
comoving domain $D$ (with the dust particles) at a fixed time $t$ is
defined as~\cite{Buchert:1999er}
\begin{eqnarray}
\langle O \rangle_D\equiv \frac{1}{V_D(t)}\int_D J(t,{\bf
x})O(t,{\bf x})\mbox{d}\bf x,\label{average}
\end{eqnarray}
where $V_D(t) \equiv \int_D J(t,{\bf x}) \mbox{d}{\bf x}$ is the
volume of the domain, and the boundary of the domain is assumed to
be comoving. Following this spatial averaging procedure, we
calculate the averaged expansion rate $\langle \theta\rangle_D$ as
an example.

From the definition of $V_D(t)$, we may introduce an effective scale
factor $a_D$
\begin{eqnarray}
\frac{a_D}{a_{D_0}}\equiv
\left(\frac{V_D(t)}{V_{D_0}}\right)^{1/3},\label{adefinition}
\end{eqnarray}
where $a_{D_0}$ and $V_{D_0}$ are the values of $a_{D}$ and $V_D$ at
the present time. With the help of (\ref{jtheta}) and
(\ref{adefinition}) we find the averaged expansion rate $\langle
\theta\rangle_D$
\begin{eqnarray}
\langle \theta\rangle_D=\frac{1}{V_D}\int_D \theta J \mbox{d}{\bf
x}=\frac{1}{V_D}\int_D \dot{J} \mbox{d}{\bf x}=
\frac{\dot{V}_D}{V_D}=3\frac{\dot{a}_D}{a_D}.\label{thetaaver}
\end{eqnarray}
The effective Hubble expansion rate can thus be defined as
\begin{eqnarray}
H_D\equiv \frac{\dot{a}_D}{a_D}=\frac{1}{3}\langle
\theta\rangle_D.\label{h}
\end{eqnarray}
The expansion rate $\langle \theta\rangle_D$ and the Hubble
expansion rate $H_D$ naturally reduce to $3 H$ and $H \equiv
\dot{a}/a$ in the homogeneous and isotropic case.

An important consequence of the definition (\ref{average}) is that
the spatial average and the time derivative do not commute with each
other. It is straightforward to prove a corresponding Lemma
(commutation rule)~\cite{Buchert:1999er}
\begin{eqnarray}
\langle O\rangle^{^{\textbf{.}}}_D-\langle \dot{O}\rangle_D= \langle
O\theta\rangle_D-\langle O\rangle_D\langle
\theta\rangle_D.\label{lemma}
\end{eqnarray}
This Lemma is used to calculate the second order term of the
averaged expansion rate $\langle \theta\rangle_D$ in Section 6.

\subsection{Buchert equations} \label{sec:bucherteqs}

With the definition of the spatial average (\ref{average}) and the
Lemma (\ref{lemma}), we yield the Buchert
equations~\cite{Buchert:1999er} from averaging the Einstein
equations (\ref{adm00}) -- (\ref{admij}) and the Raychaudhuri
equation (\ref{ray}),
\begin{eqnarray}
&&\left(\frac{\dot{a}_D}{a_D}\right)^2 + \frac{k_D}{a_D^2}=
\frac{8\pi G}{3}\rho_{\rm eff}, \label{b1} \\
&&-\frac{\ddot{a}_D}{a_D} = \frac{4\pi G}{3}(\rho_{\rm eff}+3p_{\rm
eff}), \label{b2}
\end{eqnarray}
where $\rho_{\rm eff}$ and $p_{\rm eff}$ are the effective energy
density and effective pressure of an isotropic fluid, which read
\begin{eqnarray}
&&\rho_{\rm eff} \equiv \langle \rho\rangle_D-\frac{1}{16\pi G}
\left[ \langle Q\rangle_D + \left(\langle R \rangle_D -
\frac{6 k_D}{a_D^2}\right)\right], \label{rhoeff} \\
&&p_{\rm eff} \equiv - \frac{1}{16\pi G}\left[\langle Q\rangle_D-
\frac{1}{3}\left(\langle R
\rangle_D-\frac{6k_D}{a_D^2}\right)\right]. \label{peff}
\end{eqnarray}
The expression $\langle Q \rangle_D$ is the kinematical backreaction
term,
\begin{eqnarray}
\langle Q\rangle_D &\equiv& \frac{2}{3}\langle (\theta-\langle
\theta\rangle_D)^2\rangle_D-2\langle \sigma^2\rangle_D \n\\
&=&\frac{2}{3}(\langle \theta^2\rangle_D-\langle \theta\rangle_D^2)
-2\langle \sigma^2\rangle_D, \label{qd}
\end{eqnarray}
which consists of the variance of the averaged expansion rate and
the averaged shear scalar.

Equations (\ref{b1}) and (\ref{b2}) express a highly nontrivial
result! They closely resemble the Friedmann equations, but have been
obtained without the assumption of homogeneity and isotropy. What
has been shown is that \emph{any} irrotational dust Universe,
averaged over comoving (spatial) domains appears to the observers to
be a FLRW-like Universe. The ``curvature term'' $k_D/a_D^2$ has been
introduced to show that actually any FLRW geometry might be picked.
Two geometries differ by their expressions for the effective fluid
(\ref{rhoeff}) and (\ref{peff}). Without loss of generality, we can
thus take $k_D = 0$ in the following calculations.

This formulation of backreaction provides the link to arguments in
favor of cosmological backreaction that have been put forward by one
of the authors in~\cite{Schwarz:2002ba}. It has been argued that on
the largest scales (where the cosmic principle applies) we can view
the Universe as being described by a FLRW model filled with a single
isotropic, but imperfect fluid, i.e., we can then understand
structure formation as an dissipative process that creates entropy
and it has been shown in~\cite{Schwarz:2002ba} that the second law
of thermodynamics implies for an expanding dust Universe that
$p_{\rm eff} < 0$.

From the Buchert equations, we see that the evolution of the
inhomogeneous and anisotropic Universe depends not only on the
energy density, but also the backreaction term $\langle Q \rangle_D$
and the averaged spatial curvature $\langle R \rangle_D$. So it is
quite important to know the values of $\langle Q\rangle_D$ and
$\langle R \rangle_D$. For instance, we find from (\ref{b2}) that if
$\rho_{\rm eff}+3p_{\rm eff}<0$, i.e., $\langle Q\rangle_D>4\pi
G\langle \rho\rangle_D$, the averaged expansion accelerates. In
other words, the averaged Universe can expand in an accelerating way
in the dust era, even if the local expansion is decelerating
everywhere in the Universe. \emph{Accelerated expansion of the
averaged expansion rate does not violate the strong energy
condition!}

We calculate $\langle Q \rangle_D$ and $\langle R\rangle_D$ in
cosmological perturbation theory to both first and second orders in
the next two sections. Furthermore, we can define the effective
equation of state as
\begin{eqnarray}
w_{\rm eff}\equiv \frac{p_{\rm eff}}{\rho_{\rm eff}}=\frac{\langle
R\rangle_D-3\langle Q\rangle_D}{2\langle \theta\rangle_D^2}
,\label{w}
\end{eqnarray}
and the square of an effective speed of sound as
\begin{eqnarray}
c_{\rm eff}^2 \equiv \frac{\dot{p}_{\rm eff}}{\dot{\rho}_{\rm
eff}}.\label{speed}
\end{eqnarray}
This effective speed of sound is the characteristic speed at which a
small perturbation propagates through the effective fluid. An
example would be a deformation of the boundary of the domain, or a
perturbation caused by the introduction of some extra mass into the
domain. Our effective speed of sound is different from the
isentropic speed of sound. We calculate $w_{\rm eff}$ and $c_{\rm
eff}^2$ in Section 6.

\subsection{Integrability condition} \label{sec:integrability}

The Buchert equations contain two averaged quantities, $\langle Q
\rangle_D$ and $\langle R \rangle_D$, which influence the evolution
of the inhomogeneous and anisotropic Universe. However, these two
terms are not independent, but can be linked by an integrability
condition.

In the irrotational dust universe, pressure is negligible, so the
energy-momentum tensor is given by $T^{\mu}_{\nu}=\rho
u^{\mu}u_{\nu}$. From the covariant conservation of its time-like
part we find the continuity equation
\begin{eqnarray}
\dot{\rho}=-\theta \rho. \label{thetarho}
\end{eqnarray}
The space-like conservation is the Euler equation, which is trivial
for an irrotational dust Universe, if expressed in the comoving
synchronous coordinates. Taking the average of (\ref{thetarho}) and
applying the Lemma (\ref{lemma}) we have
\begin{equation}
\langle \rho \rangle_D^{^{\textbf{.}}} = -\langle \theta \rangle_D
\langle \rho\rangle_D = -3H_D\langle \rho\rangle_D. \label{rhodot}
\end{equation}

From (\ref{b1}), (\ref{b2}) and (\ref{rhodot}), we find the relation
between $\langle Q \rangle_D$ and $\langle R\rangle_D$ (the
integrability condition)~\cite{Buchert:1999er}
\begin{eqnarray}
(a_D^6\langle Q\rangle_D)^{^{\textbf{.}}}+a_D^4(a_D^2\langle
R\rangle_D)^{^{\textbf{.}}}=0.\label{int}
\end{eqnarray}
The integrability condition is an essential equation for the
following calculations. Its advantage is that it can be used to any
order in perturbative calculations, as it is an exact result. This
is shown in Section 6, where we make use of the integrability
condition to derive the second order terms of $\langle R \rangle_D$,
$\langle \theta \rangle_D$ and $\langle \rho \rangle_D$ without
using the metric perturbations of second order. If we had not made
the choice $k_D = 0$, we would need to replace $\langle R\rangle_D$
in (\ref{int}) by $\langle R\rangle_D - 6 k_D/a_D^2$, which would
not change the solution of the integrability condition, as can be
seen easily.

So far, all our results are exact for an irrotational dust universe.
In order to get quantitative information on the observed Universe,
we turn to cosmological perturbation theory. We use the comoving
synchronous gauge below.

\section{Linearized Einstein equations in the comoving synchronous gauge}
\label{sec:pert}

In this section, we first introduce the metric perturbations $\Psi$
and $\chi$ and find the linearized Einstein equations. Solving these
equations, we find the time dependence of $\Psi$ and $\chi$. With
the help of these solutions, we calculate $\langle Q \rangle_D$,
$\langle R \rangle_D$, $\langle \theta \rangle_D$ and $\langle \rho
\rangle_D$ to both first and second orders in the next two sections.

\subsection{Einstein equations for the perturbed Universe} \label{sec:einstein}

We start now from a spatially flat FLRW dust model with the scale
factor $a(t)$. In the synchronous gauge we write the first order
linearly perturbed metric as~\cite{Kolb:2004am}
\begin{eqnarray}
\mbox{d}s^2=-\mbox{d}t^2+a^2(t)[(1-2\Psi)\delta_{ij}+D_{ij}\chi]\mbox{d}x^i
\mbox{d}x^j,\label{metric}
\end{eqnarray}
where $\Psi$ and $\chi$ are the scalar metric perturbations at first
order, $D_{ij} \equiv
\partial_{i}\partial_{j}-\frac{1}{3}\delta_{ij}\Delta$ and $\Delta$
denotes the Laplace operator in a three-dimensional Euclidean space.
The scale factor $a$ in (\ref{metric}) is not the same as the
effective scale factor $a_D$ defined in (\ref{adefinition}), and
their relation is shown in Section 6.

From the line element (\ref{metric}), we straightforwardly obtain
the nontrivial Christoffel symbols,
\begin{eqnarray}
\Gamma^0_{ij}&=&a\dot{a}\delta_{ij}-2a\dot{a}\Psi\delta_{ij}-a^2\dot{\Psi}\delta_{ij}
+a\dot{a}D_{ij}\chi+\frac{a^2}{2}D_{ij}\dot{\chi},  \nonumber\\
\Gamma^i_{0j}&=&\frac{\dot{a}}{a}\delta^i_{j}-\dot{\Psi}\delta^i_{j}
+\frac{1}{2}D^i_j\dot{\chi},\nonumber\\
\Gamma^i_{jk}&=&-\partial_k\Psi\delta^i_{j}-\partial_j\Psi\delta^i_{k}+\partial^i\Psi\delta_{jk}
\n\\
&&+\frac{1}{2}D^i_j\partial_k\chi+\frac{1}{2}D^i_k\partial_j\chi-\frac{1}{2}D_{jk}\partial^i\chi,
\label{gammapert}
\end{eqnarray}
and the components of the Einstein tensor
\begin{eqnarray}
G^0_0&=&-3\left(\frac{\dot{a}}{a}\right)^2+6\frac{\dot{a}}{a}\dot{\Psi}
-\frac{2}{a^2}\left(\Psi+\frac{1}{6}\Delta\chi\right), \nonumber\\
G^0_i&=&-2\partial_i\left(\Psi+\frac{1}{6}\Delta\chi\right)^{\textbf{.}}, \nonumber\\
G^i_j&=&-\left[\left(\frac{\dot{a}}{a}\right)^2+2\frac{\ddot{a}}{a}\right]\delta^i_j
\n\\
&&+\frac{2}{3}\left[3\ddot{\Psi}+9\frac{\dot{a}}{a}\dot{\Psi}-\frac{1}{a^2}\Delta
\left(\Psi+\frac{1}{6}\Delta\chi\right)\right]\delta^i_j\nonumber\\
&&+D^i_j\left[\frac{1}{2}\ddot{\chi}+\frac{3\dot{a}}{2a}\dot{\chi}
+\frac{1}{a^2}\left(\Psi+\frac{1}{6}\Delta\chi\right)\right].\n
\end{eqnarray}
The energy-momentum tensor of dust becomes
\begin{eqnarray}
&&T^{0}_{0}=-\rho=-\rho^{(0)}-\rho^{(1)},
\end{eqnarray}
where $\rho^{(0)}$ and $\rho^{(1)}$ are the energy density of the
background and at first order, respectively.

We are now ready to obtain the linearized equations of motion in the
ADM decomposition. The different components at different orders are \\
the energy constraint at  zeroth order
\begin{eqnarray}
\left(\frac{\dot{a}}{a}\right)^2=\frac{8\pi
G}{3}\rho^{(0)},\label{000}
\end{eqnarray}
and at first order
\begin{eqnarray}
- 3\frac{\dot{a}}{a}\dot{\Psi} +
\frac{1}{a^2}\Delta\left(\Psi+\frac{1}{6}\Delta\chi\right) = 4\pi
G\rho^{(1)}; \label{001}
\end{eqnarray}
the momentum constraint
\begin{eqnarray}
\partial_i\left(\Psi+\frac{1}{6}\Delta\chi\right)^{\textbf{.}}=0;\label{0i}
\end{eqnarray}
the evolution equation at zeroth order
\begin{eqnarray}
\left(\frac{\dot{a}}{a}\right)^2+2\frac{\ddot{a}}{a}=0,\label{ij0}
\end{eqnarray}
its diagonal ($i=j$) piece at first order
\begin{eqnarray}
3\ddot{\Psi}+9\frac{\dot{a}}{a}\dot{\Psi}
-\frac{1}{a^2}\Delta\left(\Psi+\frac{1}{6}\Delta\chi\right)=0,\label{ij}
\end{eqnarray}
and its off-diagonal ($i\neq j$) piece at first order
\begin{eqnarray}
D^i_j\left[\frac{1}{2}\ddot{\chi}+\frac{3\dot{a}}{2a}\dot{\chi}
+\frac{1}{a^2}\left(\Psi+\frac{1}{6}\Delta\chi\right)
\right]=0.\label{ji}
\end{eqnarray}

From the covariant energy-momentum conservation, we find at zeroth
order
\begin{eqnarray}
\dot{\rho}^{(0)}+3\frac{\dot{a}}{a}\rho^{(0)}=0,\label{em0}
\end{eqnarray}
and at first order
\begin{eqnarray}
\dot{\rho}^{(1)}+3\frac{\dot{a}}{a}\rho^{(1)}-3\dot{\Psi}\rho^{(0)}=0.\label{em1}
\end{eqnarray}
Equation (\ref{em1}) has a first integral,
\begin{equation}
\bar\zeta ({\bf x}) \equiv \frac{\rho^{(1)}}{3 \rho^{(0)}} - \Psi,
\label{zeta}
\end{equation}
which resembles the famous hypersurface-invariant variable (here for
dust, expressed in the synchronous coordinates)
\begin{equation}
\zeta(t, {\bf x})  \equiv  \bar\zeta - \frac 16 \Delta \chi,
\end{equation}
commonly used to characterize the primordial power spectrum
~\cite{Bardeen:1988hy}.

\subsection{Solutions for $a$, $\Psi$ and $\chi$} \label{sec:psichi}

\paragraph{Solution for $a$.}
From (\ref{em0}), we have $\rho^{(0)}a^3=\rho^{(0)}_0a^3_0$, where
$\rho^{(0)}_0$ and $a_0$ are the values of $\rho^{(0)}$ and $a$ at
the present time. By means of (\ref{000}), we find
\begin{eqnarray}
\frac{a}{a_0}=\left(\frac{t}{t_0}\right)^{2/3}.\label{a}
\end{eqnarray}
We can see that $a$ grows as $t^{2/3}$, which is the result of the
spatially flat FLRW dust cosmology. But this does not mean that the
perturbed Universe expands in the same way as the unperturbed one,
because in the perturbed Universe, the meaningful scale factor is
the effective scale factor $a_D$ defined in (\ref{adefinition}),
which, however, is not equivalent to $a$, and their relation is
obtained in Section 6. So we cannot know the behavior of the
expansion of the perturbed Universe in terms of the scale factor
$a$.

\vspace{0.3cm}

\paragraph{Solution for $\Psi$.}

We first eliminate $\rho^{(1)}$ from the equations of motion with
the help of the first integral $\bar\zeta$ from (\ref{zeta}),
\begin{eqnarray}
\rho^{(1)} = \frac{3\rho^{(0)}_0a^3_0}{a^3} \left(\Psi +
\bar\zeta({\bf x})\right). \label{rho1}
\end{eqnarray}
This allows us to obtain an equation for $\Psi$. Namely, from
(\ref{001}), (\ref{ij}) and (\ref{rho1}), we have
\begin{eqnarray}
3\ddot{\Psi}+6\frac{\dot{a}}{a}\dot{\Psi}= 4\pi G \rho^{(1)} =
\frac{12 \pi G \rho^{(0)}_0 a^3_0}{a^3}\left(\Psi + \bar\zeta({\bf
x})\right), \label{phi}
\end{eqnarray}
and using (\ref{a}), we obtain
\begin{eqnarray}
3\ddot{\Psi}+\frac{4}{t}\dot{\Psi}-\frac{2}{t^2}\Psi=\frac{2}{t^2}
\bar\zeta({\bf x}).\n
\end{eqnarray}
So we find the solution for $\Psi$
\begin{eqnarray}
\Psi({\bf x}, t) = A({\bf x})t^{2/3}+B({\bf x})t^{-1}-\bar\zeta({\bf
x}),\label{psi}
\end{eqnarray}
where $A({\bf x})$ and $B({\bf x})$ are constants of integration,
i.e., functions of the spatial coordinates only. We can see from
(\ref{psi}) that $\Psi$ consists of one growing mode $A({\bf
x})t^{2/3}$, one decaying mode $B({\bf x})t^{-1}$, and one constant
mode $-\bar\zeta({\bf x})$, in which the free spatial functions must
be fixed by the initial conditions.

In the next sections, we see that only the time derivatives
$\dot{\Psi}$ and $\ddot{\Psi}$ show up in the observables that are
of interest to us. Also, we are only concerned with the evolutions
of perturbations at late times, so we can neglect the decaying mode
$B({\bf x})t^{-1}$. Hence, only the growing mode $A({\bf x})t^{2/3}$
is of importance for the following calculations. Thus, we get the
time derivatives of $\Psi$
\begin{eqnarray}
\dot{\Psi}=\frac{2}{3}A({\bf x})t^{-1/3}, \quad
\ddot{\Psi}=-\frac{2}{9}A({\bf x})t^{-4/3}.\label{psipert1}
\end{eqnarray}

\vspace{0.3cm}

\paragraph{Solution for $\chi$.}

From $G^i_j=0$, we have
\begin{eqnarray}
&&\frac{2}{3}\left[3\ddot{\Psi}+9\frac{\dot{a}}{a}\dot{\Psi}-\frac{1}{a^2}\Delta
\left(\Psi+\frac{1}{6}\Delta\chi\right)\right]\delta^i_j\n\\
&&+D^i_j\left[\frac{1}{2}\ddot{\chi}+\frac{3\dot{a}}{2a}\dot{\chi}
+\frac{1}{a^2}\left(\Psi+\frac{1}{6}\Delta\chi\right)\right]=0.
\label{gij}
\end{eqnarray}
Multiplying $a^2$ on (\ref{gij}), taking the time derivative and
inserting (\ref{0i}), we have
\begin{eqnarray}
\left(\frac{a^2}{2}D^i_j\ddot{\chi}+\frac{3a\dot{a}}{2}D^i_j\dot{\chi}\right)^{^{\textbf{.}}}
+\left[a^2\left(2\ddot{\Psi}
+6\frac{\dot{a}}{a}\dot{\Psi}\right)\right]^{^{\textbf{.}}}\delta^i_j=0.\label{0}
\end{eqnarray}
From (\ref{a}) and (\ref{psipert1}), the second part in (\ref{0}) is
0, so we get
\begin{eqnarray}
D^i_j\stackrel{...}{\chi}+\frac{10}{3t}D^i_j\ddot{\chi}+\frac{2}{3t^2}D^i_j\dot{\chi}=0,\n
\end{eqnarray}
and thus
\begin{eqnarray}
D^i_j\dot{\chi}=C^i_j({\bf x})t^{-1/3}+E^i_j({\bf x})t^{-2},\n
\end{eqnarray}
where $C^i_j({\bf x})$ and $E^i_j({\bf x})$ are functions of spatial
coordinates only. In the following calculation we neglect the
decaying mode $E^i_j({\bf x})t^{-2}$, so
\begin{eqnarray}
D^i_j\dot{\chi}=C^i_j({\bf x})t^{-1/3}. \label{chidot}
\end{eqnarray}
From (\ref{chidot}), we obtain the solution for $\chi$
\begin{eqnarray}
\chi({\bf x}, t)=C({\bf x})t^{2/3}+g({\bf x}),\label{chi}
\end{eqnarray}
where $C({\bf x})$ and $g({\bf x})$ are also the functions of
spatial coordinates only. The function $g({\bf x})$ does not carry
physical information, as it can be fixed by the residual spatial
gauge transformation of the comoving synchronous gauge, and we
utilize this freedom to set $g ({\bf x})= 0$. Substituting
({\ref{chi}}) into ({\ref{chidot}}), we can get the relation between
$C({\bf x})$ and $C^i_j({\bf x})$
\begin{eqnarray}
C^i_j({\bf x})=\frac{2}{3}D^i_j C({\bf x}).\nonumber
\end{eqnarray}
Finally, the time derivatives of $\chi$ are
\begin{eqnarray}
\dot{\chi}=\frac{2}{3}C({\bf x})t^{-1/3},\quad
\ddot{\chi}=-\frac{2}{9}C({\bf x})t^{-4/3}.\label{chipert1}
\end{eqnarray}

We see from (\ref{psi}) and (\ref{chi}) that both $\Psi$ and $\chi$
grow as $t^{2/3}$ at late times. Because $a\propto t^{2/3}$, both
$\Psi$ and $\chi$ grow linearly as the scale factor $a$ in the
perturbed dust Universe. So if cosmological perturbation theory is
valid, i.e., the perturbative terms $\Psi$ and $\chi$ are small, the
scale factor should not be too large. In other words, a perturbative
analysis (at any order) is restricted to the ``linear'' regime.

\subsection{Relation between $A({\bf x})$, $C({\bf x})$ and $\bar\zeta({\bf
x})$}\label{sec:aczeta}

We have obtained the solutions for $\Psi$ and $\chi$, but these two
solutions are not independent due to (\ref{0i}). From (\ref{0i}),
(\ref{psipert1}) and (\ref{chipert1}), we have
\begin{eqnarray}
\partial_i\left(A+\frac{1}{6}\Delta C\right)=0.\nonumber
\end{eqnarray}
Because both $A$ and $C$ are functions of the spatial coordinates
only, we have
\begin{eqnarray}
A+\frac{1}{6}\Delta C=K,\label{ack}
\end{eqnarray}
where $K$ is a constant. In the spatially flat universe $K=0$, which
is shown in the next section, and we find the relation between $A$
and $C$, i.e., the relation between $\Psi$ and $\chi$
\begin{eqnarray}
A=-\frac{1}{6}\Delta C.\label{ac}
\end{eqnarray}
Finally, with help of the evolution equations (\ref{ij}) and
(\ref{ji}), we obtain
\begin{equation}
\bar \zeta({\bf x}) = \frac{5}{9} \frac{a_0^2}{t_0^{4/3}} C({\bf
x}).
\end{equation}
Let us note that at superhorizon scales $\bar \zeta \approx \zeta$.
As the amplitude of $\zeta$ at superhorizon scales is measured by
cosmic microwave background experiments, the magnitudes of the time
independent functions $\bar \zeta$, $C$ and $A$ are thus fixed.

\subsection{Relation between $C({\bf x})$, $\bar\zeta({\bf x})$ and the peculiar gravitational
potential $\varphi({\bf x})$}\label{sec:czetaphi}

The peculiar gravitational potential $\varphi({\bf x})$ is defined
from the Poisson equation as~\cite{Kolb:2004am}
\begin{eqnarray}
\Delta\varphi({\bf x})\equiv 4\pi G\rho^{(1)}a^2.\label{ccphi}
\end{eqnarray}
From (\ref{phi}), using (\ref{psipert1}) and (\ref{ac}), we have
\begin{eqnarray}
\Delta C({\bf x})=-12\pi G\rho^{(1)}t^{4/3}.\label{cphi}
\end{eqnarray}
So, with the help of (\ref{ccphi}), (\ref{cphi}) and (\ref{a}), we
obtain the relation between $C({\bf x})$, $\bar\zeta({\bf x})$ and
$\varphi({\bf x})$
\begin{eqnarray}
\varphi({\bf x})=-\frac{1}{3}\frac{a_0^2}{t_0^{4/3}}C({\bf x}) = -
\frac {3}{5} \bar\zeta({\bf x}). \label{phic}
\end{eqnarray}
We see that the peculiar gravitational potential $\varphi({\bf x})$
is just a linear function of the metric perturbation $\chi$.

Therefore, we know the solutions for $a$, $\Psi$ and $\chi$, which
we use in the next two sections to calculate the first and second
order contributions to the averaged physical quantities, and we
focus on the investigation of the time dependence of the averaged
observables.

\section{First order perturbations} \label{sec:first}

In this section, we calculate $\langle \theta\rangle_D$, $\langle
R\rangle_D$ and $\langle \rho\rangle_D$ in the perturbed Universe,
and these results are the first step to the derivation of the second
order contributions. We do not calculate $\langle Q \rangle_D$,
because it is a pure second order term, which is proven in the next
section.

For the first order calculations of averaged quantities, the
integration measure $J$ must be expanded to the first order as well,
\begin{eqnarray}
J=a^3(1-3\Psi)=a^3\left(1-3A t^{2/3}\right)
=a^3\left(1+\frac{1}{2}\Delta Ct^{2/3}\right),\n
\end{eqnarray}
as at late times, the decaying and constant modes are negligible. In
the following, let us denote
\begin{equation}
\langle O \rangle_{D1} \equiv \frac{\int_D O\mbox{d}\bf x}{\int_D
\mbox{d}\bf x},
\end{equation}
which is defined to be the average on the background (i.e., $J=
a^3$). Watch out that the average is still over a physically
comoving domain, which might have a distorted geometry, even on the
background. Thus, for the first order calculations, the averages of
the zeroth and first order quantities are
\begin{eqnarray}
&&\langle O^{(0)} \rangle_D = \frac{\int_D O^{(0)}J\mbox{d}\bf
x}{\int_D J\mbox{d}\bf x}=O^{(0)},\n\\
&&\langle O^{(1)} \rangle_D = \frac{\int_D O^{(1)}J\mbox{d}\bf
x}{\int_D J\mbox{d}\bf x}=\frac{\int_D O^{(1)}\mbox{d}\bf x}{\int_D
\mbox{d}\bf x}= \langle O^{(1)} \rangle_{D1}.\n\\\label{o1}
\end{eqnarray}
\emph{Therefore, the perturbation in $J$ does not affect the first
order calculations.}

\subsection{Averaged expansion rate $\langle \theta \rangle_D$}
\label{sec:theta1}

Substituting the perturbative connections in (\ref{gammapert}) into
(\ref{theta}), we have
\begin{eqnarray}
\theta^i_j= \frac{\dot{a}}{a}\delta^i_j-\dot{\Psi}\delta^i_j
+\frac{1}{2}D^i_j\dot{\chi},\label{thetaijpert1}
\end{eqnarray}
and taking the trace, we get the perturbative expansion rate to
first order,
\begin{eqnarray}
\theta=3\frac{\dot{a}}{a}-3\dot{\Psi},\label{thetapert1}
\end{eqnarray}
in which we have used the property $D^i_i=0$. Using (\ref{a}) and
(\ref{psipert1}), we obtain the averaged expansion rate $\langle
\theta \rangle_D$ as a function of cosmic time $t$ and $A$,
\begin{eqnarray}
\langle \theta\rangle_D=3\frac{\dot{a}}{a}-3\langle
\dot{\Psi}\rangle_D =\frac{2}{t}-\frac{2\langle
A\rangle_{D1}}{t^{1/3}}.\label{thetapert1aver}
\end{eqnarray}
From (\ref{thetapert1aver}), the first order perturbation decays as
$t^{-1/3}$, which is slower than that of the zeroth order term
$2/t$. Therefore, the perturbation becomes more and more important
as the Universe evolves. However, this does not mean that this
perturbation dominates at late times, as in the perturbative
approach, we must restrict our analysis to $|\langle \Psi
\rangle_D|\ll 1$.

\subsection{Averaged spatial curvature $\langle R \rangle_D$} \label{sec:r1}

From (\ref{adm00}) and the trace of (\ref{admij}), we have
\begin{eqnarray}
&&R=16\pi G\rho-\theta^2+\theta^i_j\theta^j_i,\n\\
&&R=12\pi G\rho-\dot{\theta}-\theta^2,\label{adm00ii}
\end{eqnarray}
and thus
\begin{eqnarray}
R=-\theta^2-4\dot{\theta}-3\theta^i_j\theta^j_i.\label{rpert1}
\end{eqnarray}
By means of (\ref{thetaijpert1}), (\ref{thetapert1}), (\ref{a}) and
(\ref{psipert1}), we find to first order
\begin{eqnarray}
\langle R\rangle_D= \frac{40\langle
A\rangle_{D1}}{3t^{4/3}}.\label{rpert1aver}
\end{eqnarray}

Closer inspection of (\ref{rpert1aver}) shows that
\begin{enumerate}
\item $\langle R\rangle_D$ has only a first order term
while the zeroth order term vanishes, as the background metric is
spatially flat.

\item $\langle R\rangle_D$ decays as $t^{-4/3}$. From (\ref{a}),
$a\propto t^{2/3}$, so $\langle R\rangle_D \propto 1/a^2$ as the
Universe expands~\cite{Rasanen:2006kp}.

\item we may, with the help of (\ref{ack}), rewrite $\langle R\rangle_D$ as
$\langle
R\rangle_D=\frac{40}{3t^{4/3}}\left(-\frac{1}{6}\langle\Delta
C\rangle_{D1}+K\right)$. So we can see that the constant $K$ in
(\ref{ack}) contributes to the averaged spatial curvature a term
$40K/3t^{4/3}\propto 1/a^2$. We know that in the unperturbed $k\neq
0$ universe, $R=6k/a^2$. Thus, $40K/3t^{4/3}$ is just the background
spatial curvature term, and as we discuss the perturbations in the
spatially flat Universe, this term vanish naturally. This is the
reason for $A=-\Delta C/6$ in (\ref{ac}).
\end{enumerate}

\subsection{Averaged energy density $\langle \rho \rangle_D$} \label{sec:rho1}

Similarly, from (\ref{adm00ii}), (\ref{thetaijpert1}) and
(\ref{thetapert1}), we have
\begin{eqnarray}
\langle \rho\rangle_D=-\frac{\langle \dot{\theta}\rangle_{D}+\langle
\theta^i_j\theta^j_i\rangle_{D}}{4\pi G} =\frac{1}{6\pi
Gt^2}+\frac{\langle A\rangle_{D1}}{2\pi
Gt^{4/3}}.\label{rhopert1aver}
\end{eqnarray}
So we find for a domain overdense in average that $\langle
A\rangle_{D1}$ is positive. At the same time, from
(\ref{thetapert1aver}) the averaged expansion rate is reduced. This
is consistent with the intuitional understanding of the
gravitational collapse, which decreases the expansion rate of the
Universe. Also, from (\ref{rpert1aver}) we have a positive averaged
spatial curvature for the overdense regions.

We find in the first order perturbative calculations that only
$\Psi$ enters the expressions of $\langle \theta \rangle_D$,
$\langle R \rangle_D$ and $\langle \rho \rangle_D$, and the metric
perturbation $\chi$ does not show up. We show in the next section
that $\sigma^2=\frac{1}{8}D^i_j\dot{\chi}D^j_i\dot{\chi}$, so $\chi$
is related to the shear of the perturbed Universe. This means that
only the expansion influences the evolution of the averaged spatial
curvature term and the averaged energy density in the perturbed
Universe at first order.

Let us finally note that {\it the first order contributions to
$\langle \theta \rangle_D$, $\langle R \rangle_D$ and $\langle \rho
\rangle_D$ are all surface terms actually}, as we may write them as
integrals of total derivatives
\begin{eqnarray}
\langle A \rangle_{D1}= - \frac{\int_D \partial^i (\partial_i C)
{\rm d}{\bf x}}{6\int_D {\rm d}{\bf x}}.\n
\end{eqnarray}
More surface terms show up below, when we turn to the second order
perturbations.

\section{Second order perturbations} \label{sec:second}

We move on to the second order perturbations of physical quantities.
Second order cosmological perturbation theory has been discussed
widely in the literature, such
as~\cite{Acquaviva:2002ud,Matarrese:1997ay,Bartolo:2003bz,Kolb:2004am}.
However, in these papers, the metric perturbations of second order
are always needed for calculations, and these calculations are
always rather complicated and tedious. In this paper, we show how to
obtain the leading terms of second order contributions to $\langle Q
\rangle_D$, $\langle R \rangle_D$, $\langle \theta \rangle_D$ and
$\langle \rho \rangle_D$ from the metric perturbations of first
order only.

We first prove that the kinematical backreaction term $\langle Q
\rangle_D$ is a second order term, and then use the integrability
condition, which is a crucial new input, to find the second order
terms of $\langle R \rangle_D$, $\langle \theta \rangle_D$ and
$\langle \rho \rangle_D$. In these calculations, the shear
$\sigma^2$ and thus $\dot{\chi}$ show up in the expressions. The
effective equation of state $w_{\rm eff}$ and the square of the
effective speed of sound $c^2_{\rm eff}$ are also given to second
order.

Different from the first order contributions, at second order we
have to consider the perturbation of the integration measure $J$.
Therefore, the averaged quantities of physical observables of
different orders become
\begin{eqnarray}
\langle O^{(0)} \rangle_D &=& O^{(0)},\n\\
\langle O^{(1)} \rangle_D &=&
   \langle O^{(1)} \rangle_{D1}-3\langle O^{(1)}\Psi \rangle_{D1}
   +3\langle O^{(1)} \rangle_{D1}\langle \Psi \rangle_{D1},\n\\
\langle O^{(2)} \rangle_D &=& \langle O^{(2)} \rangle_{D1}.
\label{o2}
\end{eqnarray}
We can see that at second order, {\it the average of a first order
quantity $\langle O^{(1)} \rangle_D$ picks up two second order
modifications $-3\langle O^{(1)}\Psi \rangle_{D1}+3\langle O^{(1)}
\rangle_{D1}\langle \Psi \rangle_{D1}$}. In the following, we show
that these modifications show up in the second order calculation of
$\langle \theta \rangle_D$.

\begin{widetext}
\subsection{Averaged kinematical backreaction term $\langle Q \rangle_D$} \label{sec:q2}

Let us recall the kinematical backreaction term $\langle Q
\rangle_D$ defined in (\ref{qd})
\begin{eqnarray}
\langle Q\rangle_D\equiv \frac{2}{3}\left(\langle
\theta^2\rangle_D-\langle \theta\rangle_D^2\right) -2\langle
\sigma^2\rangle_D.\nonumber
\end{eqnarray}
Now we prove that $\langle Q \rangle_D$ is a pure second order term.
We show that the first part $\langle \theta^2\rangle_D-\langle
\theta\rangle_D^2$ and the second part $\langle \sigma^2\rangle_D$
are both of second order.

To calculate the variance $\langle \theta^2\rangle_D-\langle
\theta\rangle_D^2$ to second order, we write
\begin{eqnarray}
\theta=\theta^{(0)}+\theta^{(1)}+\theta^{(2)},
\end{eqnarray}
where $\theta^{(0)}$, $\theta^{(1)}$ and $\theta^{(2)}$ are the
zeroth, first and second order contributions to $\theta$,
respectively. $\theta^{(0)}$ and $\theta^{(1)}$ have been calculated
in (\ref{thetapert1}). Using (\ref{o2}), to second order we have
\begin{eqnarray}
\langle \theta^2\rangle_D-\langle \theta\rangle_D^2=\langle
(\theta^{(0)}+\theta^{(1)}+\theta^{(2)})^2\rangle_D-(\langle
\theta^{(0)}+\theta^{(1)}+\theta^{(2)}\rangle_D)^2=\langle
(\theta^{(1)})^2\rangle_D-\langle \theta^{(1)}\rangle_D^2=\langle
(\theta^{(1)})^2\rangle_{D1}-\langle \theta^{(1)}\rangle_{D1}^2.
\end{eqnarray}
This means that the first part of $\langle Q \rangle_D$ is of second
order, however, for calculating it we do not need to know the
detailed form of $\theta^{(2)}$. All we need is $\theta$ up to first
order (see (\ref{thetapert1})).

Similarly, we calculate the average of the shear scalar $\langle
\sigma^2\rangle_D$. From (\ref{sigma}) and (\ref{thetaijpert1}) we
find at first order that
\begin{eqnarray}
\sigma^i_j=\theta^i_j-\frac{1}{3}\theta\delta^i_j
=\theta_j^{i(0)}+\theta_j^{i(1)}-\frac{1}{3}(\theta^{(0)}+\theta^{(1)})\delta^i_j
=\frac{1}{2}D^i_j\dot{\chi},
\end{eqnarray}
so $\sigma^i_j$ has no zeroth order contribution. Hence, using
(\ref{chipert1}), we have
\begin{eqnarray}
\sigma^2=\frac{1}{2}\sigma^i_j\sigma^j_i=\frac{1}{8}D^i_j\dot{\chi}D^j_i\dot{\chi}
=\frac{1}{18t^{2/3}}\left[\partial^i\partial_jC\partial^j\partial_iC
-\frac{1}{3}(\Delta C)^2\right].\label{sigma2pert}
\end{eqnarray}
This means that $\langle \sigma^2\rangle_D$, the second part of
$\langle Q \rangle_D$), is also of second order, and can be
calculated by using the expression of $\chi$ only at first order.

So far, we have proved that both two parts of $\langle Q \rangle_D$
are of second order, and consequently $\langle Q \rangle_D$ is a
second order term, but nevertheless can be calculated from the first
order contributions to $\Psi$ and $\chi$. Using (\ref{sigma2pert}),
(\ref{thetapert1}), (\ref{psipert1}) and (\ref{chipert1}), we get
$\langle Q \rangle_D$ at second order
\begin{eqnarray}
\langle Q\rangle_D=6\left(\langle \dot{\Psi}^2\rangle_{D1}-\langle
\dot{\Psi}\rangle_{D1}^2\right)-\frac{1}{4}\langle
D^i_j\dot{\chi}D^j_i\dot{\chi} \rangle_{D1}
=\frac{1}{27t^{2/3}}\left[3\left(\langle (\Delta
C)^2\rangle_{D1}-\langle
\partial^i\partial_jC\partial^j\partial_iC\rangle_{D1}\right)
-2\langle \Delta C\rangle_{D1}^2\right]
\equiv\frac{F}{t^{2/3}},\label{qpertaver2}
\end{eqnarray}
where
\begin{eqnarray}
F\equiv \frac{1}{27}\left[3\left(\langle (\Delta
C)^2\rangle_{D1}-\langle
\partial^i\partial_jC\partial^j\partial_iC\rangle_{D1}\right)
-2\langle \Delta C\rangle_{D1}^2\right]
=\frac{1}{27}\left[3\left(\langle
\partial^i(\partial_iC\Delta C)\rangle_{D1} - \langle
\partial^i(\partial_jC\partial^j\partial_iC)\rangle_{D1}\right) -
2\langle\Delta C\rangle_{D1}^2\right].\nonumber
\end{eqnarray}
$F$ has only second order terms and is a function of spatial
coordinates, and here we use (\ref{ac}) to express the final result
in terms of the variable $C$ only.

Some remarks on this result for $\langle Q \rangle_D$ are in order.
From (\ref{qpertaver2}) we find that
\begin{enumerate}
\item $\langle Q \rangle_D$, written in the form abbreviated $F$,
contains two second order terms, which are total derivatives and
become surface terms when averaging. Meanwhile, the third term
$\langle\Delta C\rangle_{D1}^2$ is the square of a first order
surface term, and thus its second order modifications in (\ref{o2})
do not show up in $F$. Therefore, the kinematical backreaction is a
function of the derivatives of $C$ on the boundary of the averaged
domain only.

\item Because $\Delta C$ is a fluctuating term, and can be stochastically
positive in some regions and negative in others, its average is
expected to be negligible, if the averaged domains become larger
(but are still on subhorizon scales). However,
$\partial^i\partial_jC\partial^j\partial_iC$ and $(\Delta C)^2$ are
positive definite, and therefore give nontrivial surface terms when
averaging. Thus, $\langle Q \rangle_D$ consists of these two surface
terms on large scales. If they cancel, we can say that there is no
kinematical backreaction at second order. In Newtonian limit, this
cancelation was discussed in~\cite{Buchert:1995fz} for periodic
boundary conditions, and in~\cite{Sicka:1999cb,Buchert:1999pq} for
spherically symmetric spaces. In relativistic cosmological
perturbation theory, this problem was treated
in~\cite{Rasanen:2003fy,Kasai:2006bt}. However, in general case,
there is no reason for this cancelation. A review of this
cancelation problem can be found in~\cite{Rasanen:2006kp}.

\item We can see from (\ref{qpertaver2}) that $\langle Q \rangle_D$
decreases as $t^{-2/3}$, which means that $\langle Q\rangle_D\propto
1/a$. And we already know that $\langle \rho^{(0)}\rangle_D\propto
1/a^3$ and $\langle R\rangle_D\propto 1/a^2$. So the kinematical
backreaction term $\langle Q \rangle_D$ decays slower than $\langle
\rho\rangle_D$ and $\langle R\rangle_D$. Therefore, in the course of
the expansion of the Universe the kinematical backreaction becomes
more and more important in the effective energy density $\rho_{\rm
eff}$ and effective pressure $p_{\rm eff}$.  Of course, we should
pay attention that $\langle Q \rangle_D$ is a pure second order
term, but $\langle R\rangle_D$ has got a first order term, and
$\langle \rho\rangle_D$ even contains a zeroth order term, so we
cannot conclude that $\langle Q \rangle_D$ dominates the late time
evolution of the Universe. The effect of the kinematical
backreaction on the expansion of the perturbed Universe depends on
its evolution with time $t$ (or scale factor $a$) in its
denominator, and also on the value of the surface terms in the
numerator.
\end{enumerate}

\subsection{Averaged spatial curvature $\langle R \rangle_D$} \label{sec:r2}

In the last section, we have calculated the averaged spatial
curvature $\langle R \rangle_D$ to first order in
(\ref{rpert1aver}), and in this subsection, we use the integrability
condition (\ref{int}) to get its second order part. But to do so, we
first need to find the relation between the effective scale factor
$a_D$, which is defined as the cubic root of the volume of integral
in the averaging procedure in (\ref{adefinition}), and the scale
factor $a$, which is defined to describe the expansion of the
Universe in the perturbative metric (\ref{metric}), and then express
$a_D$ as the function of the perturbation $A$. Since the
integrability condition is an exact relation to any order, and we
have already got $\langle Q \rangle_D$ to second order in
(\ref{qpertaver2}), solving the integrability condition
$(a_D^6\langle Q\rangle_D)^{^{\textbf{.}}}+a_D^4(a_D^2\langle
R\rangle_D)^{^{\textbf{.}}}=0$, it is straightforward to obtain
$\langle R \rangle_D$ to second order. However, in the following, we
show that we do not need to calculate $a_D$ to second order, but
only first order is sufficient for our purpose.

We may rewrite the integrability condition as
\begin{eqnarray}
6\frac{\dot{a}_D}{a_D}\langle Q\rangle_D+\langle
Q\rangle_D^{^{\textbf{.}}}+ 2\frac{\dot{a}_D}{a_D}\langle
R\rangle_D+\langle R\rangle_D^{^{\textbf{.}}}=0.\label{intexp}
\end{eqnarray}
Because $\langle Q \rangle_D$ is already of second order, in the
first term of (\ref{intexp}), we only need the zeroth order term of
$\dot{a}_D/a_D$. In the third term, since $\langle R\rangle_D$ has
no zeroth order term, we need the zeroth and first order terms of
$\dot{a}_D/a_D$.

From (\ref{thetaaver}) and (\ref{thetapert1}), to first order we
have
\begin{eqnarray}
\frac{\dot{a}_D}{a_D}=\frac{1}{3}\langle
\theta\rangle_{D1}=\frac{\dot{a}}{a}-\langle
\dot{\Psi}\rangle_{D1}=\frac{2}{3t}-\frac{2\langle
A\rangle_{D1}}{3t^{1/3}}, \label{ad}
\end{eqnarray}
so at first order
\begin{eqnarray}
\frac{a_D(t)}{a_D(t_0)}&=&\left(\frac{t}{t_0}\right)^{2/3}
\left[1-\langle A\rangle_{D1}\left(t^{2/3}-t^{2/3}_{0}
\right)\right].\label{adaver1}
\end{eqnarray}
Thus, if $\langle A\rangle_{D1}$ is negative, the effective scale
factor $a_D$ grows faster than the ordinary result $t^{2/3}$ in the
unperturbed dust model. We can also see this from
(\ref{rhopert1aver}) that if $\langle A\rangle_{D1}<0$, $\langle
\rho\rangle_D$ is reduced with respect to the background and the
expansion rate gets positive modification, which is consistent with
the intuition that the underdense regions expand faster than the
overdense ones.

Substituting (\ref{ad}) and (\ref{qpertaver2}) into (\ref{intexp}),
to second order we have
\begin{eqnarray}
\langle
R\rangle_D^{^{\textbf{.}}}+\frac{4}{3}\left(\frac{1}{t}-\frac{\langle
A\rangle_{D1}}{t^{1/3}}\right)\langle
R\rangle_D+\frac{10}{3}\frac{F}{t^{5/3}}=0.\n
\end{eqnarray}
Solving this differential equation, we find
\begin{eqnarray}
\langle R\rangle_D=D\left(\frac{t_0}{t}\right)^{4/3}
\mbox{exp}\left(2\langle
A\rangle_{D1}\left(t^{2/3}-t^{2/3}_0\right)\right)
+\frac{5}{2}\frac{F}{\langle A\rangle_{D1}t^{4/3}}
\left[1-\mbox{exp}\left(2\langle
A\rangle_{D1}\left(t^{2/3}-t^{2/3}_0\right)\right)\right],\label{rsolution}
\end{eqnarray}
where $D$ is the constant of integration, which is a function of
spatial coordinates. For consistency, we must expand the solution up
to second order
\begin{eqnarray}
\langle R\rangle_D=\frac{1}{t^{4/3}}\left(Dt^{4/3}_0-2\langle
A\rangle_{D1}Dt^{2}_0+5Ft^{2/3}_0\right)+\frac{1}{t^{2/3}}\left(2\langle
A\rangle_{D1}Dt^{4/3}_0-5F\right).\n
\end{eqnarray}
There is only one undetermined constant of integration $D$ in the
above expression. From (\ref{rpert1aver}), we know that $\langle
R\rangle_D$ has no zeroth order term, so $D$ has only the first and
second order terms, otherwise, the terms in the first bracket would
give rise to a zeroth order contribution. We may write
$D=D^{(1)}+D^{(2)}$, where $D^{(1)}$ and $D^{(2)}$ are the first and
second order terms of $D$, respectively. Because $\langle
A\rangle_{D1}$ is a first order term and $F$ is a second order one,
$\langle R\rangle_D$ can be written as
\begin{eqnarray}
\langle
R\rangle_D=\frac{1}{t^{4/3}}(D^{(1)}t^{4/3}_0+D^{(2)}t^{4/3}_0-2\langle
A\rangle_{D1}D^{(1)}t^{2}_0+5Ft^{2/3}_0)+\frac{1}{t^{2/3}}(2\langle
A\rangle_{D1}D^{(1)}t^{4/3}_0-5F),\label{rpertaver2}
\end{eqnarray}
where the first term $D^{(1)}t^{4/3}_0/t^{4/3}$ represents the first
order term of $\langle R\rangle_D$. It is matched with the first
order expression (\ref{rpert1aver}) to fix
\begin{eqnarray}
D^{(1)}=\frac{40\langle A\rangle_{D1}}{3t_0^{4/3}}.\label{d}
\end{eqnarray}
Substituting (\ref{d}) into (\ref{rpertaver2}), and using
(\ref{ac}), we find the averaged spatial curvature $\langle
R\rangle_D$ to second order,
\begin{eqnarray}
\langle R\rangle_D=-\frac{20}{9t^{4/3}}\langle \Delta
C\rangle_{D1}+\frac{G^{(2)}}{t^{4/3}}+\frac{5}{9t^{2/3}}\left[\left(\langle
\partial^i(\partial_jC\partial^j\partial_iC)\rangle_{D1}
-\langle \partial^i(\partial_iC\Delta C)\rangle_{D1}\right)+2\langle
\Delta C\rangle_{D1}^2\right],\label{rpert2}
\end{eqnarray}
where $G^{(2)}\equiv D^{(2)}t^{4/3}_0-2\langle
A\rangle_{D1}D^{(1)}t^{2}_0+5Ft^{2/3}_0$. Neither $D^{(2)}$ nor
$G^{(2)}$ can be fixed by matching to some known coefficients.
However, the term $G^{(2)}/t^{4/3}$ is unimportant at any time.
Early on, $-\frac{20}{9}\langle \Delta C\rangle_{D1}$ is a first
order term, while $G^{(2)}$ is a second order one, so it is
negligible compared to the first term in (\ref{rpert2}). Similarly,
at late times, $G^{(2)}/t^{4/3}$ decays faster than
$\frac{5}{9t^{2/3}}\left[\langle
\partial^i(\partial_jC\partial^j\partial_iC)\rangle_{D1}
-\langle \partial^i(\partial_iC\Delta C)\rangle_{D1}+2\langle \Delta
C\rangle_{D1}^2\right]$, because both numerators are of second
order, but the exponent of the denominator in $G^{(2)}/t^{4/3}$ is
the larger one. Thus, $-\frac{20}{9t^{4/3}}\langle \Delta
C\rangle_{D1}$ is the first order term of $\langle R\rangle_D$,
which is the same as the result in (\ref{rpert1aver}), and
$\frac{5}{9t^{2/3}}\left[\langle
\partial^i(\partial_jC\partial^j\partial_iC)\rangle_{D1}
-\langle \partial^i(\partial_iC\Delta C)\rangle_{D1}+2\langle \Delta
C\rangle_{D1}^2\right]$ is the \emph{leading} second order part at
late times. Therefore, in the following calculations, we write
$\langle R\rangle_D$ as
\begin{eqnarray}
\langle R\rangle_D=-\frac{20}{9t^{4/3}}\langle \Delta
C\rangle_{D1}+\frac{5}{9t^{2/3}}\left[\left(\langle
\partial^i(\partial_jC\partial^j\partial_iC)\rangle_{D1}
-\langle \partial^i(\partial_iC\Delta C)\rangle_{D1}\right)+2\langle
\Delta C\rangle_{D1}^2\right].\label{rpertnog2}
\end{eqnarray}
Thus, at second order, $\langle R\rangle_D$ is again the function of
surface terms. So with (\ref{rpert1aver}), we find that $\langle
R\rangle_D$ is the function of surface terms at both first and
second orders.

In this subsection, we have extended the calculation of the averaged
spatial curvature $\langle R\rangle_D$ to second order by using the
integrability condition. Its advantage is that we can do the second
order calculation, without knowing the metric perturbations of
second order. This is because the integrability condition is an
exact result to any order, and we have got $\langle Q \rangle_D$ to
second order with only the first order perturbation theory.

\subsection{Averaged expansion rate $\langle \theta \rangle_D$}
\label{sec:theta2}

The second order perturbation of the expansion rate has been
discussed in the literature. For instance, in~\cite{Kolb:2004am},
Kolb {\it et al.} used the metric perturbations of second order to
calculate the averaged Hubble expansion rate and its variance. Here,
the expansion rate $\theta$ is defined in the same way as that
in~\cite{Kolb:2004am}, namely $\theta\equiv u^{\lambda}_{;\lambda}$.
However, the Hubble expansion rate in~\cite{Kolb:2004am} is defined
as $\sqrt{8\pi G \langle \rho\rangle_D/3}$, in contrast to the one
in our work, which is defined in (\ref{h}) as $H_D \equiv
\frac{\dot{a}_D}{a_D}=\frac{1}{3}\langle \theta\rangle_D$. Below,
the second order perturbation of the expansion rate is calculated,
but without using the metric perturbations of second order again. We
also show that our calculation is consistent with the result
in~\cite{Kolb:2004am}.

From (\ref{rpert1}), we have
\begin{eqnarray}
\langle R \rangle_D= -\langle \theta^2 \rangle_D- 4\langle
\dot{\theta} \rangle_D- 3\langle \theta^i_j\theta^j_i \rangle_D= -
2\langle \theta^2 \rangle_D- 4\langle \dot{\theta} \rangle_D-
6\langle \sigma^2 \rangle_D. \label{rtheta}
\end{eqnarray}
Since $\langle R \rangle_D$ has already been calculated to second
order in (\ref{rpertnog2}), $\langle \sigma^2 \rangle_D$ is a pure
second order term, and we know the zeroth and first order terms of
$\langle \theta\rangle_D$ from (\ref{thetapert1aver}), we can obtain
the second order perturbation of $\langle \theta\rangle_D$ from
(\ref{rtheta}).

Using (\ref{thetapert1}), we expand $\theta$ as
\begin{eqnarray}
\theta=\theta^{(0)}+\theta^{(1)}+\theta^{(2)}=3\frac{\dot{a}}{a}-3\dot{\Psi}+\theta^{(2)}
=\frac{2}{t}-\frac{2A}{t^{1/3}}+\theta^{(2)}.\label{thetapert2}
\end{eqnarray}
so to second order
\begin{eqnarray}
\theta^2=\frac{4}{t^2}-\frac{8A}{t^{4/3}}
+\frac{4A^2}{t^{2/3}}+\frac{4\theta^{(2)}}{t}, \quad
\dot{\theta}=-\frac{2}{t^2}
+\frac{2}{3}\frac{A}{t^{4/3}}+\dot{\theta}^{(2)}.\label{thetadotpert2}
\end{eqnarray}
Substituting (\ref{thetadotpert2}) into (\ref{rtheta}), and using
(\ref{sigma2pert}) and (\ref{ac}), we find
\begin{eqnarray}
R=\frac{40A}{3t^{4/3}}-4\dot{\theta}^{(2)}-\frac{8}{t}
\theta^{(2)}-\frac{8 A^{2}}{t^{2/3}}-\frac{3}{4}
D^i_j\dot{\chi}D^j_i\dot{\chi}.\label{r2}
\end{eqnarray}
We see from (\ref{r2}) that $R$ has both the first and second order
terms, so at second order, the first order term
$\frac{40A}{3t^{4/3}}$ gives two additional second order
modifications when averaging as shown in (\ref{o2}). Therefore, the
average of the spatial curvature $R$ to second order is
\begin{eqnarray}
\langle R\rangle_D&=&\frac{40\langle A\rangle_D}{3t^{4/3}} -4\langle
\dot{\theta}^{(2)}\rangle_D-\frac{8}{t}\langle
\theta^{(2)}\rangle_D-\frac{8\langle
A^{2}\rangle_D}{t^{2/3}}-\frac{3}{4}\langle
D^i_j\dot{\chi}D^j_i\dot{\chi}\rangle_D\nonumber\\
&=&-\frac{20}{9t^{4/3}}\langle \Delta C\rangle_{D1}-4\langle
\theta^{(2)}\rangle_{D1}^{^{\textbf{.}}}-\frac{8}{t}\langle
\theta^{(2)}\rangle_{D1}-\frac{1}{9t^{2/3}}\left[11\langle (\Delta
C)^2\rangle_{D1} -10\langle (\Delta C)\rangle_{D1}^2 +3\langle
\partial^i\partial_jC\partial^j\partial_iC\rangle_{D1}\right].\label{theta2}
\end{eqnarray}
Above, the Lemma (\ref{lemma}) allows us to write $\langle
\dot{\theta}^{(2)}\rangle_{D1}=\langle
\theta^{(2)}\rangle_{D1}^{^{\textbf{.}}}$ at second order. Matching
({\ref{theta2}}) with (\ref{rpert2}) yields
\begin{eqnarray}
\langle \theta^{(2)}\rangle_{D1}^{^{\textbf{.}}}+\frac{2}{t}\langle
\theta^{(2)}\rangle_{D1} +\frac{1}{18t^{2/3}}\left[3\langle (\Delta
C)^2\rangle_{D1}+4\langle
\partial^i\partial_jC\partial^j\partial_iC\rangle_{D1}\right]=0.\nonumber
\end{eqnarray}
Solving this differential equation provides us with the second order
contribution to the averaged expansion rate $\langle
\theta\rangle_D$
\begin{eqnarray}
\langle \theta^{(2)}\rangle_{D1}=-\frac{t^{1/3}}{42}\left[3\langle
(\Delta C)^2\rangle_{D1}+4\langle
\partial^i\partial_jC\partial^j\partial_iC\rangle_{D1}\right]+\frac{I}{t^2},
\end{eqnarray}
where $I$ is the constant of integration, and at late times the term
$I/t^2$ is negligible without doubt. Therefore, we find the averaged
expansion rate $\langle \theta\rangle_D$ to second order,
\begin{eqnarray}
\langle \theta\rangle_D&=&\langle \theta^{(0)}\rangle_D+\langle
\theta^{(1)}\rangle_D+\langle \theta^{(2)}\rangle_D\n\\
&=&\frac{2}{t}+\frac{1}{3t^{1/3}}\langle \Delta
C\rangle_{D1}-\frac{t^{1/3}}{42}\left[4\left(\langle
\partial^i(\partial_jC\partial^j\partial_iC)\rangle_{D1}
-\langle \partial^i(\partial_iC\Delta C)\rangle_{D1}\right)
+7\langle (\Delta C)\rangle_{D1}^2 \right],\label{theta2pert2aver}
\end{eqnarray}
where the first order term $\theta^{(1)}$ also contributes via
averaging to the second order result (see (\ref{o2})).
Straightforwardly, from (\ref{rhodot}), the averaged Hubble
expansion rate is
\begin{eqnarray}
H_D=\frac{2}{3t}+\frac{1}{9t^{1/3}}\langle \Delta
C\rangle_{D1}-\frac{t^{1/3}}{126}\left[4\left(\langle
\partial^i(\partial_jC\partial^j\partial_iC)\rangle_{D1}
-\langle \partial^i(\partial_iC\Delta C)\rangle_{D1}\right)
+7\langle (\Delta C)\rangle_{D1}^2 \right].\nonumber
\end{eqnarray}
We can see that $\langle \theta\rangle_D$ and $\langle H\rangle_D$
are also functions of surface terms at both first and second orders.

Finally, we show that the result in (\ref{theta2pert2aver}) can also
be obtained by using the metric perturbations of second order
in~\cite{Kolb:2004am},
\begin{equation}
\mbox{d}s^2=a^2(\eta)\left[-\mbox{
d}\eta^2+\left(\left(1-2\Psi^{(1)}-\Psi^{(2)}\right)\delta_{ij}
+D_{ij}\left(\chi^{(1)}+\frac{1}{2}\chi^{(2)}\right)
+\frac{1}{2}\left(\partial_i \chi^{(2)}_j+\partial_j
\chi^{(2)}_i+\chi^{(2)}_{ij}\right)\right)
\mbox{d}x^i\mbox{d}x^j\right],\label{kolb}
\end{equation}
where $\eta$ is the conformal time, $\Psi^{(1)}$ and $\chi^{(1)}$
are the first order scalar perturbations (the same as in this
paper), $\Psi^{(2)}$ and $\chi^{(2)}$ are the second order scalar
perturbations, $\chi_i^{(2)}$ is the second order transverse vector
perturbation, and $\chi_{ij}^{(2)}$ is the second order transverse
and traceless tensor perturbation. Ignoring decaying modes, for a
dust Universe, $\Psi^{(1)}$, $\chi^{(1)}$ and $\Psi^{(2)}$ are also
given in~\cite{Kolb:2004am},
\begin{eqnarray}
\Psi^{(1)}(\eta, {\bf x})&=&\frac{3}{5}\varphi({\bf
x})+\frac{\eta^2}{18}\Delta\varphi({\bf x}), \quad
\chi^{(1)}(\eta, {\bf x})=-\frac{\eta^2}{3}\varphi({\bf x}),\n\\
\Psi^{(2)}(\eta, {\bf x})&=&-\frac{50}{9}\varphi^2({\bf
x})-\frac{5\eta^2}{54}\partial^i\varphi({\bf
x})\partial_i\varphi({\bf
x})+\frac{\eta^4}{252}\left[(\Delta\varphi({\bf
x}))^2-\frac{10}{3}\partial^i\partial_j\varphi({\bf
x})\partial^j\partial_i\varphi({\bf x})\right].\label{kolbpert}
\end{eqnarray}
From (\ref{kolb}) and (\ref{kolbpert}), in the comoving synchronous
gauge, we have
\begin{eqnarray}
\theta=u^{\lambda}_{;\lambda}=\frac{1}{a}\Gamma^i_{0i}=\frac{3a'}{a^2}
-\frac{3}{a}\Psi^{(1)'}-\frac{3}{2a}\Psi^{(2)'}-\frac{6}{a}\Psi^{(1)}\Psi^{(1)'}
-\frac{1}{2a}D^{ij}\chi^{(1)}D_{ij}\chi^{(1)'},\n
\end{eqnarray}
where $'$ denotes the derivative with respect to the conformal time
$\eta$, and in the dust Universe, $t=(a_0\eta)^3/(27t_0^2)$. So
using (\ref{phic}), we can straightforwardly get the average of the
expansion rate $\theta$,
\begin{eqnarray}
\langle \theta\rangle_D&=&\frac{2}{t}+\frac{1}{3t^{1/3}}\langle
\Delta C\rangle_{D1}-\frac{5a^2_0}{54t^{4/3}_0t^{1/3}}\left[\langle
\partial^i(C\partial_iC)\rangle_{D1}+\langle C\Delta C\rangle_{D1}
-6\langle C\rangle_{D1}\langle \Delta C\rangle_{D1}\right]\n\\
&&-\frac{t^{1/3}}{42}\left[4\left(\langle
\partial^i(\partial_jC\partial^j\partial_iC)\rangle_{D1}
-\langle \partial^i(\partial_iC\Delta C)\rangle_{D1}\right)
+7\langle (\Delta C)\rangle_{D1}^2 \right].\label{kolbtheta}
\end{eqnarray}
Thus, we find that the leading second order term in
(\ref{kolbtheta}), which we get by using the metric perturbations of
second order in~\cite{Kolb:2004am}, is the same as that in
(\ref{theta2pert2aver}). One can see as already argued for the case
of $\langle R\rangle_D$ that the subleading second order
contributions show the same time dependence as the first order term.
Thus, it is justified to neglect the subleading terms as they can
never (in the perturbative regime) overcome the first order ones.

\subsection{Averaged energy density $\langle \rho \rangle_D$} \label{sec:rho2}

Similarly, from (\ref{thetaaver}) and (\ref{b1}), we have
\begin{eqnarray}
\left(\frac{1}{3}\langle \theta\rangle_D\right)^2=\frac{8\pi
G}{3}\left(\langle \rho\rangle_D-\frac{\langle Q\rangle_D+\langle
R\rangle_D}{16\pi G}\right).
\end{eqnarray}
Using (\ref{qpertaver2}), (\ref{rpertnog2}) and
(\ref{theta2pert2aver}), we get the averaged energy density to
second order,
\begin{eqnarray}
\langle \rho\rangle_D=\frac{1}{6\pi
Gt^{2}}\left[1-\frac{t^{2/3}}{2}\langle \Delta C\rangle_{D1}
+\frac{t^{4/3}}{28}\left(2\left(\langle
\partial^i(\partial_jC\partial^j\partial_iC)\rangle_{D1}
-\langle(\partial^i(\partial_iC\Delta C)\rangle_{D1}\right)+7\langle
\Delta C\rangle_{D1}^2 \right)\right],\label{rhopert2aver}
\end{eqnarray}
and $\langle \rho\rangle_D$ is a function of surface terms at both
first and second orders too.

Up to this point, we have obtained all the averaged quantities
$\langle Q\rangle_D$, $\langle R\rangle_D$, $\langle
\theta\rangle_D$ and $\langle \rho\rangle_D$ to second order, and we
only have to use the first order perturbation terms $\Psi$ and
$\chi$ (without the necessity of knowing the metric perturbations of
second order). These simplifications are based on the integrability
condition and the fact that $\langle Q \rangle_D$ is a pure second
order term. Unfortunately, we cannot extend this method to higher
orders. For example, if we go to third order, to calculate
$\sigma^2$, we would need $\theta^i_j=\Gamma^i_{0j}$ to second
order, and thus we must know the metric perturbations of second
order. An exception is the irrotational and shearless universe,
which is merely the FLRW model.

\subsection{Effective equation of state and square of effective speed of sound} \label{sec:es}

{\it a. Effective equation of state $w_{\rm eff}$.} From (\ref{w}),
using (\ref{qpertaver2}), (\ref{rpert2}) and (\ref{rhopert2aver}),
we obtain the effective equation of state to second order,
\begin{eqnarray}
w_{\rm eff}=w^{(1)}_Dt^{2/3}+w^{(2)}_Dt^{4/3} =-\frac{5}{18}\langle
\Delta C\rangle_{D1}t^{2/3}+\left[\frac{1}{9}\left(\langle
\partial^i(\partial_jC\partial^j\partial_iC)\rangle_{D1}
-\langle(\partial^i(\partial_iC\Delta
C)\rangle_{D1}\right)+\frac{7}{27}\langle \Delta
C\rangle_{D1}^2\right]t^{4/3}.\label{wpertaver2}
\end{eqnarray}
Equation (\ref{wpertaver2}) is the perturbative expansion of the
effective equation of state, which is both time and domain
dependent. We can find its time dependence at different orders, and
domain dependence in different coefficients. From (\ref{ad}), it is
easy to get $t$ as the function of the effective scale factor $a_D$,
and at late times (i.e., $t\gg t_0$) $w_{\rm eff}$ can be rewritten
as
\begin{eqnarray}
w_{\rm eff}&=&w'^{(1)}_Da_D+w'^{(2)}_Da_D^2 \n\\
&=&-\frac{5\langle\Delta
C\rangle_{D1}}{18}\frac{t^{2/3}_0}{a_{D_0}}a_D+\left[\frac{1}{9}\left(\langle
\partial^i(\partial_jC\partial^j\partial_iC)\rangle_{D1}
-\langle(\partial^i(\partial_iC\Delta
C)\rangle_{D1}\right)+\frac{11}{36}\langle \Delta
C\rangle_{D1}^2\right]\frac{t^{4/3}_0}{a^2_{D_0}}a_D^2.\label{dj}
\end{eqnarray}
Equation (\ref{dj}) is also a perturbative expansion of the
effective equation of state, but in terms of $a_D$, which is of more
interest than $a$. We can see that the second order coefficients in
(\ref{wpertaver2}) and (\ref{dj}) are different, this is because
$a_D$ is not proportional to $t^{2/3}$, and thus the second order
coefficient picks up nontrivial contributions from the first order
one.

Therefore, $w_{\rm eff}$ vanishes at zeroth order. This is different
from the cosmological constant, for which $w_{\Lambda}=-1$.
Consequently, in a perturbative framework, the backreaction
mechanism cannot induce accelerated expansion of the Universe as
that would imply $w_{\rm eff} < -1/3$. Nevertheless, the
cosmological perturbations allow us to investigate a possible change
of the expansion rate of the averaged Universe that might in the
later nonlinear stage lead to the accelerated expansion of the
Universe. We discuss this on both small and large scales.

Using (\ref{cphi}), we may rewrite $w_{\rm eff}$ as
\begin{eqnarray}
w_{\rm eff}=\frac{10\pi G}{3}\langle
\rho^{(1)}\rangle_{D1}t^{2}+\left[\frac{1}{9}\left(\langle
\partial^i(\partial_jC\partial^j\partial_iC)\rangle_{D1}
-\langle(\partial^i(\partial_iC\Delta
C)\rangle_{D1}\right)+\frac{7}{27}\langle \Delta
C\rangle_{D1}^2\right]t^{4/3}. \label{wpert2aver}
\end{eqnarray}

Firstly, on small scales, $\langle \rho^{(1)}\rangle_{D1}$ may
significantly deviate from 0, so the first order term dominates the
value of $w_{\rm eff}$, i.e., $w_{\rm eff}=\frac{10\pi G}{3}\langle
\rho^{(1)}\rangle_{D1}t^{2}$. We can see from (\ref{wpert2aver})
that if $\langle \rho^{(1)}\rangle_{D1}<0$, which means that the
energy density is underdense locally, $w_{\rm eff}$ is negative, and
since $w_{\rm eff}\propto t^2$, this effect will be more and more
influential as time goes on, and might cause the accelerated
(averaged) expansion of the inhomogeneous and anisotropic Universe.
Of course, with the above expression we can trace the evolution only
for small perturbations. Once they are in the nonlinear regime, our
approach fails.

Secondly, for large averaged domains (but are still on subhorizon
scales), like the discussion on $F$, the average of $\Delta C$ is
expected to become negligible, since it is a fluctuating term, and
only the two surface terms give nontrivial contributions. Therefore
the value of $w_{\rm eff}$ is dominated by these two second order
terms on large scales,
\begin{eqnarray}
w_{\rm eff}=\frac{1}{9}\left(\langle
\partial^i(\partial_jC\partial^j\partial_iC)\rangle_{D1}-\langle
\partial^i(\partial_iC\Delta C)\rangle_{D1}\right)t^{4/3}.\label{weff}
\end{eqnarray}
We can see from (\ref{weff}) that the sign of $w_{\rm eff}$ depends
on the contrast of the two surface terms. It vanishes for certain
boundary conditions,
see~\cite{Buchert:1995fz,Sicka:1999cb,Buchert:1999pq,
Rasanen:2003fy,Kasai:2006bt,Rasanen:2006kp}. However, we think that
these boundary conditions are not natural and that the generic case
for a finite domain in the Universe is that the effective equation
of state is given by a finite surface term, that might be positive
or negative, depending on the details of the fluctuations on the
boundaries.

An important lesson that we learn here is that {\it the cosmological
backreaction introduces an effective equation of state, which is not
only time dependent, but also scale dependent}.

\vspace{0.3cm}

{\it b. Square of the effective speed of sound $c_{\rm eff}^2$.}
Similarly, for the square of the effective speed of sound, we have
\begin{eqnarray}
c_{\rm eff}^2=-\frac{5}{27}\langle \Delta
C\rangle_{D1}t^{2/3}+\frac{1}{27}\left[\left(\langle
\partial^i(\partial_jC\partial^j\partial_iC)\rangle_{D1}
-\langle(\partial^i(\partial_iC\Delta
C)\rangle_{D1}\right)+\frac{47}{18}\langle \Delta
C\rangle_{D1}^2\right]t^{4/3}. \label{cpertaver2}
\end{eqnarray}

Like $w_{\rm eff}$, on small scales,
\begin{eqnarray}
c_{\rm eff}^2=-\frac{5}{27}\langle \Delta
C\rangle_{D1}t^{2/3}=\frac{20\pi G}{9}\langle
\rho^{(1)}\rangle_{D1}t^2.\nonumber
\end{eqnarray}
So if the cosmic medium is overdense locally, $c_{\rm eff}^2>0$. But
we also see that $c_{\rm eff}^2$ can be negative in underdense
regions. Usually this suggest that some damping is going on, which
is related to dissipative phenomena and the increase of entropy.
These aspects will be investigated in more detail elsewhere.

On large scales, the second order terms dominate and we find
\begin{eqnarray}
c_{\rm eff}^2&=&\frac{1}{27}\left(\langle
\partial^i(\partial_jC\partial^j\partial_iC)\rangle_{D1}-\langle
\partial^i(\partial_iC\Delta C)\rangle_{D1}\right)t^{4/3}.\n
\end{eqnarray}
Also the sign of the square of the effective speed of sound depends
on the contrast of the two surface terms.

To summarize this section, we can see that \emph{all studied
physical quantities, $\langle Q \rangle_D$, $\langle R\rangle_D$,
$\langle \theta \rangle_D$, $H_D$, $\rho_{\rm eff}$, $w_{\rm eff}$
and $c_{\rm eff}^2$, can be expressed as functions of surface terms
at both first and second orders}. Thus, to know the values of these
physical quantities, we do not need to know anything about the
interior of the averaged domain. Only the physical information
(i.e., the values of $C$, namely the peculiar gravitational
potential $\varphi$, and its derivatives) encoded on the boundary of
the domain matters.
\end{widetext}

\section{Gauge dependence of the averaged quantities}

We should finally discuss the gauge dependence of the averaged
physical observables. In~\cite{Bruni:1996im,Matarrese:1997ay}, the
gauge invariance of physical observables at different orders are
discussed in detail: to second order, a quantity is gauge dependent
unless its zeroth and first order terms vanish, apart from the
trivial cases that it is a constant scalar field, or a linear
combination of products of Kronecker deltas with constant
coefficients on the background. Thus, we know that $\langle Q
\rangle_D$ is a gauge invariant quantity, since it has only the
second order term. Secondly, $\langle R\rangle_D$, $w_{\rm eff}$ and
$c_{\rm eff}^2$, which have the first order terms, and $\langle
\theta \rangle_D$, $H_D$ and $\rho_{\rm eff}$, which have both the
zeroth and first order terms depend on the gauge choice. However,
the first order terms of $\langle R\rangle_D$, $w_{\rm eff}$ and
$c_{\rm eff}^2$ are gauge independent as well. So in summary, we
conclude that all leading terms of all physical observables are
gauge invariant, while the higher order ones are not.

This raises the question of the coordinate system. It seems to us
that the comoving synchronous gauge is very close to the coordinate
system of a real observer. Real astronomers and their telescopes are
comoving with matter (we neglect the difference between baryonic and
dark matter here), they use their own proper time in all of their
observations and regard space to be time-orthogonal, which defines
precisely the slicing and gauge that we use throughout this work.

\section{Conclusions and Discussions} \label{sec:conclusion}

In this paper, we use both the Buchert equations and cosmological
perturbation theory to study the evolution of the perturbed dust
Universe in the comoving synchronous gauge. We investigate the
possibility to explain the accelerated expansion of the Universe
without dark energy. We calculate the averaged kinematical
backreaction term $\langle Q \rangle_D$ and the averaged spatial
curvature $\langle R\rangle_D$, and find that $\langle Q \rangle_D$
is a pure second order term, and $\langle R\rangle_D$ has both the
first and second order terms. As we use a perturbative approach,
these terms can only affect the evolution of the Universe
perturbatively, and thus we can only hope to find an onset of the
cosmological backreaction mechanism in this work. In some
circumstances, for example, on small scales, on which the effects of
fluctuations of the energy density $\langle \rho\rangle_D$ are
significantly nonzero, the backreaction mechanism should not be
neglected carelessly.

We conclude that cosmological backreaction is for real and that it
can both increase or decrease the expansion of the averaged
Universe, depending on the averaged domain under consideration. Thus
we argue that the effective equation of state of the Universe is
time and scale dependent and so is the square of the effective speed
of sound.

We find in our perturbative approach, that all physical quantities
are surface terms or squares of surface terms. This suggests the
conjecture that also a nonlinear treatment would find only functions
of surface terms.

Another point of our paper is that we show in Section 6 how to
calculate the averaged quantities to second order for the leading
growing mode, but use only the metric perturbations of first order.
This is a consequence of the integrability condition, which is valid
to any order. And this greatly simplifies the perturbative
calculations, which are usually done with help of the metric
perturbations of second order in the previous papers.

Finally, we discuss some observational aspects of the perturbative
calculations in our paper. The problem is how to measure the
perturbations, such as $\Psi$ or $A$.

The fluctuation amplitude $\sigma^2_D$ is defined as
\begin{eqnarray}
\sigma^2_D&\equiv& \left\langle
\left(\frac{\delta\rho}{\rho}\right)^2\right\rangle_D=\left\langle
\left(\frac{\rho^{(1)}+\rho^{(2)}}{\rho}\right)^2\right\rangle_D \n\\
&=&\left\langle
\left(\frac{\rho^{(1)}}{\rho}\right)^2\right\rangle_{D1}.\n
\end{eqnarray}
Here we only calculate $\sigma^2_D$ to second order, so the terms
containing $\rho^{(2)}$ are negligible. (Usually this amplitude is
defined at the distance of $8~\mpc$, but we can also define it for
any domain $D$.) From (\ref{rhopert1aver}), we have
\begin{eqnarray}
\sigma^2_D=\frac{t^{4/3}}{4}\langle (\Delta C)^2\rangle_{D1}.
\end{eqnarray}
So by measuring $\sigma^2_D$, we can know the value of $\langle
(\Delta C)^2\rangle_{D1}$.

Secondly, a measurement of the averaged energy density $\langle
\rho\rangle_D$ in (\ref{rhopert2aver}), the averaged expansion rate
$\langle \theta \rangle_D$ in (\ref{theta2pert2aver}) and the
averaged spatial curvature $\langle R\rangle_D$ in
(\ref{rpertnog2}), together with the measurement of $\sigma^2_D$,
would allow us to find $\langle
\partial^i\partial_jC\partial^j\partial_iC\rangle_{D1}$. Because
all of them are physical observables (at least in principle), from
these measurements we can know the values of $\langle (\Delta
C)^2\rangle_{D1}$ and $\langle
\partial^i\partial_jC\partial^j\partial_iC\rangle_{D1}$. And
since the other averaged physical quantities $\langle Q \rangle_D$,
$w_{\rm eff}$ and $c_{\rm eff}^2$ are the functions of the former,
we can obtain the quantitative information of all these averaged
terms. This will help us to understand the evolution of the
inhomogeneous and anisotropic Universe and its relation to what we
call the background model on a much deeper level. It also
demonstrates that we do not need to retreat to a statistical
treatment of the backreaction effect, but we can try to design an
experiment to directly measure its sign and magnitude in the Milky
Way's neighborhood (e.g., the local $\sim (100 \mpc)^3$ domain).

\vspace{0.5cm}

{\bf \quad\quad\quad\quad ACKNOWLEDGEMENTS}

\vspace{0.3cm}

We are very grateful to Thomas Buchert, Aleksandar Raki\'c and Syksy
R\"as\"anen for discussions. N.L. also thanks Je-An Gu, Sabino
Matarrese, Alessio Notari, Sergey Piskarev and Qiang Wu. The work of
N.L. is supported by the DFG under grant GRK 881.

\end{document}